%% file: main.tex
\documentclass{webofc}
\usepackage[varg]{txfonts}   % Web of Conferences font
\usepackage[autostyle, english = american]{csquotes}
\usepackage{hyperref}
\usepackage{graphicx}
\usepackage{lineno}
\graphicspath{ {./images/} }
\newcommand{\C}{$^\circ$C }
\begin{document}

\title{Climate impacts of particle physics}

\author{\firstname{Kenneth} \lastname{Bloom}\inst{1}
\fnsep\thanks{Contact author, \email{kenbloom@unl.edu}} 
\and
\firstname{Veronique} \lastname{Boisvert}\inst{2}
\fnsep\thanks{Contact author, \email{Veronique.Boisvert@rhul.ac.uk}} 
\and
\firstname{Daniel} \lastname{Britzger}\inst{3}
\and
\firstname{Micah} \lastname{Buuck}\inst{4}
\and
\firstname{Astrid} \lastname{Eichhorn}\inst{5}
\and
\firstname{Michael} \lastname{Headley}\inst{6}
\and
\firstname{Kristin} \lastname{Lohwasser}\inst{7}
\and
\firstname{Petra} \lastname{Merkel}\inst{8}
%\fnsep\thanks{\email{petra@fnal.gov}} 
%%%
% ---> everyone please enter your name and affiliation
}

\institute{University of Nebraska-Lincoln, Lincoln, NE, USA 
\and
          Royal Holloway University London, United Kingdom
\and
          Max-Planck-Institute for Physics, Munich, Germany
\and
          SLAC National Accelerator Laboratory, Menlo Park, CA, USA
\and 
    CP3-Origins, University of Southern Denmark, Denmark       
\and
        Sanford Underground Research Facility (SURF), Lead, SD, USA   
\and
          University of Sheffield, United Kingdom
\and
          Fermi National Accelerator Laboratory (Fermilab), Batavia, IL, USA
          }

\abstract{
The pursuit of particle physics requires a stable and prosperous society.  Today, our society is increasingly threatened by global climate change.  Human-influenced climate change has already impacted weather patterns, and global warming will only increase unless deep reductions in emissions of CO$_2$ and other greenhouse gases are achieved.  Current and future activities in particle physics need to be considered in this context, either on the moral ground that we have a responsibility to leave a habitable planet to future generations, or on the more practical ground that, because of their scale, particle physics projects and activities will be under scrutiny for their impact on the climate.  In this white paper for the U.S. Particle Physics Community Planning Exercise (``Snowmass''), we examine several contexts in which the practice of particle physics has impacts on the climate.  These include the construction of facilities, the design and operation of particle detectors, the use of large-scale computing, and the research activities of scientists.  We offer recommendations on establishing climate-aware practices in particle physics, with the goal of reducing our impact on the climate. We invite members of the community to show their support for a sustainable particle physics field~\cite{bib:indico}.
}

\newcommand\snowmass{\begin{center}\rule[-0.2in]{\hsize}{0.01in}\\\rule{\hsize}{0.01in}\\
\vskip 0.1in Submitted to the  Proceedings of the US Community Study\\ 
on the Future of Particle Physics (Snowmass 2021)\\ 
\rule{\hsize}{0.01in}\\\rule[+0.2in]{\hsize}{0.01in} \end{center}}

\maketitle

\snowmass{}

\section{Introduction}
% KB and VB
\input{intro}

\section{Impacts of facility construction}
% KB and VB
\input{facility}

\section{Impacts of detector gases}
% Beatrice and Roberto
\input{gases}

\section{Impacts of computing}
% Micah
\input{computing}

\section{Impacts of scientists' research activities}
% Petra and all authors

\input{activities}

\section{Recommendations}
% all authors
\input{recs}

\section{Acknowledgments}
\input{Ack}

\input{references}

\end{document}

%% file: intro.tex
\label{sec:intro}
Global climate change, and how to mitigate it, is one of the most crucial issues facing humanity today~\cite{bib:UNoals}.  The Intergovernmental Panel on Climate Change (IPCC) has stated, ``It is unequivocal that human influence has warmed the atmosphere, ocean and land. Widespread and rapid changes in the atmosphere, ocean, cryosphere and biosphere have occurred''~\cite{bib:IPCC6}.  This change in climate has already affected weather patterns and climate extremes in regions across the globe, including in the U.S.  The IPCC further states that ``Global warming of 1.5\C and 2\C will be exceeded during the 21st century unless deep reductions in CO$_2$ and other greenhouse gas emissions occur in the coming decades.''  To limit the amount of warming, we will have to achieve significant reductions in CO$_2$ emissions, to the point of net-zero emissions, along with reductions in the production of other greenhouse gases (GHGs), in line with the goals of the Paris Agreement~\cite{bib:Paris}.  Each 1000 gigatons of cumulative CO$_2$ emissions is likely to lead to 0.27\C  to 0.63\C increases in global surface temperature, implying that limiting temperature increases means adhering to a ``carbon budget''.  The 6th IPCC assessment report estimates a total budget of 300 gigatons CO2e emissions for an 83\% chance to limit global warming to below 1.5 degrees Celsius \cite{bib:IPCC6}. This amounts to about 1.1~t CO2e\footnote{Carbon dioxide equivalent or CO2e means the number of metric tons of CO2 emissions with the same global warming potential as one metric ton of another greenhouse gas, and is calculated using Equation A-1 in 40 CFR Part 98~\cite{bib:co2e_def}.}   emissions per capita per year until 2050. This should be compared to current per capita per year emission rates of 14.2~tCO$_2$ in the United States~\cite{bib:pcCO2}.  Significant reductions in carbon emissions must be achieved for the future health of the planet, and measures must be taken now rather than later to begin to limit the long-term impacts.

The U.S is a crucial player in the climate agenda. It is the top producer and consumer of both oil and natural gas, it has the world’s second largest number of coal-fired power plants, and fossil fuels contributed 63\% to its overall electricity generation~\cite{CarbonBriefProfile}. On the other hand, it also has the largest nuclear and second largest renewable capacity in the world. Not only is the U.S. the second largest emitter of GHGs, but over the course of history it has cumulatively produced more than any other country~\cite{bib:histemissions}. Its citizens have emissions footprints that are roughly three times the global average. The U.S also has a large impact on climate policy. For example, the success of the 2015 Paris agreement was due in significant part to the leadership demonstrated by the U.S. at that time~\cite{Obama}. More recently, the current administration has submitted an updated Nationally Determined Contribution to cut greenhouse gas emissions 50-52\% below 2005 levels by 2030~\cite{NDC} and has pledged to achieve net-zero emissions by “no later than 2050”~\cite{netzero}.

Current and future activities in particle physics need to be considered in this context.  The pursuit of particle physics requires substantial construction projects; the consumption of electricity in the operation of accelerators, detectors, and computing; the use of GHGs in particle detectors; and in some cases significant amounts of travel.  All of these lead to the potential for particle physicists to have a carbon impact well above that of typical citizens, and thus particle physicists should be paying attention to the impacts of the discipline on the planet and seeking to reduce them.  This could be argued from a moral point of view -- that we have responsibilities to future generations to leave behind a habitable planet.  However, one could also justify it on more practical grounds.  Particle physics takes place on such a scale that future major projects, such as a new facility, will be subject to scrutiny for their climate impacts, and thus the field must be prepared for that scrutiny.  Should society decide that a price on carbon is necessary, then future projects will have to be prepared to pay that price.  It may be possible to argue that because of its societal benefits, some exception to constraints on carbon emission should be made for scientific activities.  However, even today emissions rates around the world tend to be above those necessary to avert a significant temperature rise.  Also, asking for such exceptions can lead to difficult discussions of the relative size and type of societal benefits from different areas of science, especially for fundamental research areas such as particle physics, where the ultimate benefits cannot be known, or questions about whether other sectors of the economy (e.g.\ health care or agriculture) provide similar benefits.  Instead, just as our field currently demonstrates world leadership in international cooperation towards common goals, we can also demonstrate world leadership in this critical area that impacts the future of society.

In this white paper for the U.S. Particle Physics Community Planning Exercise (``Snowmass"), we examine several contexts in which the practice of particle physics has impacts on the climate.  These include the construction of large-scale experimental facilities, the design and operation of particle detectors that make use of GHGs, the operation of computing facilities, and the common research activities of physicists, including long-distance travel.  We conclude with a set of recommendations for how we as a field can begin to reduce our impact on the climate.  Our work builds on that of the European Strategy Update, which made a recommendation on studying and significantly reducing the environmental impact of particle physics activities~\cite{bib:ESU}.  We hope that this work stimulates conversations within the U.S. particle physics community about how we can pursue the science that we love in a sustainable fashion.

%% file: facility.tex
\label{sec:facility}
A key goal of the Snowmass process is to identify promising opportunities to address the most important questions in particle physics.  We expect that these opportunities will require the construction of new, large-scale experimental facilities.  The building construction industry currently contributes 10\% of the world's total carbon emissions~\cite{bib:globalABC}.  If we assume that the electric grid is successfully de-carbonized by 2040, a goal of many climate plans (for example the U.S. ``has set a goal to reach 100 percent carbon pollution-free
electricity by 2035"\cite{NDC}), then construction, rather than operations, may well dominate the climate impact of a new particle physics facility.  Here we consider the climate impact of the construction of a new accelerator facility and put it in its global context.

One potential new energy-frontier accelerator facility foreseen by the particle physics community is the Future Circular Collider (FCC)~\cite{FCC}.  This accelerator would likely first collide electrons and positrons to make precision measurements and later could accelerate hadrons with a center-of-mass energy of 100 TeV, allowing for the next stage of discovery following the HL-LHC era. The FCC-ee project~\cite{bib:FCC-ee} would operate in the era of de-carbonized electricity.  The tunnel for the accelerator would be one of the longest tunnels in the world, projected at 97.75~km circumference in the conceptual design report.  In addition, 
many bypass tunnels, access shafts, large experimental caverns, and  new surface sites are planned.  It is estimated that 7~million cubic meters of spoil, a mixture of marls and sandstone, would need to be excavated.\footnote{This projected amount of excavated material is smaller than the 9~million cubic meters described in the conceptual design report, after further optimization, and possible re-use of this material is being explored~\cite{bib:miningthefuture}.}

We can attempt a rough estimate of the carbon impact of the main tunnel alone.  A bottom-up calculation is driven by the construction parameters of the tunnel.  It is expected to have an inner diameter of 5.5~m, and an excavation diameter of 6.3~m.  We assume that the region between those two diameters is filled with concrete, and that concrete is composed 15\% of cement (and we neglect the necessary steel tunnel reinforcement).  The production of cement is a significant contributor to greenhouse gas emissions, releasing a ton of CO$_2$ per ton of cement created~\cite{bib:CBcement}.\footnote{If world production of cement were a country, it would be the third largest emitter.}  We thus calculate that the main tunnel of FCC-ee alone would lead to the release of 237~ktons of CO$_2$.

Alternatively, one could make a top-down estimate of the impact using rules of thumb from studies of previous road tunnel construction~\cite{bib:tunnelCO2}.  These studies attempt to obtain a complete accounting of emissions. For example, they consider the impacts of fuel and electricity used in tunnel construction, along with those of the construction materials and of the possible release of methane inherent to the excavation process.  The carbon impacts depend very much on the ``rock mass quality" of the excavation site, with a range between 5,000 and 10,000~kg CO$_2$ per meter of tunnel length.  This leads to estimates of 489 to 978~kton of CO$_2$ for the main FCC-ee tunnel, providing an order-of-magnitude agreement with the bottom-up estimate. Using 500~kton of CO$_2$ as a conservative estimate and dividing by the rough number of about 6000 physicists that could be contributing to this project, this amounts to about 80~t of emissions per physicist, to be compared with the target of reaching 1.1~t of emissions per person per year introduced in Section~\ref{sec:intro}. Alternatively, we estimate that 6 million trees would need to be planted to absorb this amount of CO$_2$\cite{bib:trees}.
 
How does this compare to other forms of civil construction?  For context, we can compare with the carbon impact of typical buildings.  A study~\cite{bib:buildings} estimates that the ``embodied"  carbon emissions during building construction is 500-600~kg of CO2e per square meter; we use 550~kg CO2e as a working value.  As a sample building we take New York City's 1 World Trade Center, a prominent recently-built skyscraper, which is 94 stories tall and 3.5~Mft$^2$.  The embodied carbon in 1 WTC is thus 197~ktons, meaning that the FCC-ee main tunnel alone has a carbon impact several times as large as one of the most significant building projects in the U.S.\ in recent years.

When the complete FCC-ee, or any similar-scale particle physics facility, is considered -- the full tunnel system, the additional buildings on the site, the materials for the accelerator and detectors -- we expect that the project will have a carbon impact similar to that of the redevelopment of a neighborhood of a major city.  This implies that the environmental impact of a future facility is going to receive the same scrutiny as that of a major urban construction project.  Our field needs to be prepared for this scrutiny, by preparing to collect and analyze data on carbon impacts, and also for taking reasonable measures for the reduction of climate impacts through the development and use of low-carbon materials, with a prioritized use of reused and recycled materials.  We can already begin investments in R\&D on how to reduce our carbon impact to prepare for future environmental reviews.  

Some efforts on this, beyond the matter of facility construction, are already in progress within the accelerator community.  For example there is a definite focus on the energy efficiency and power reduction of future large accelerators. This ranges for example from active research on energy recovery schemes~\cite{bib:ERL, bib:Plasma}, to using the heat produced by accelerator operations for district heating~\cite{bib:districtheating}. 

Any facility construction project will have a measurable impact on climate change, and as responsible citizens we must consider what we can do to reduce the potential harms to future generations by our actions.

%% file: gases.tex
\begin{figure}[htbp!]
\centering
\includegraphics[width=12cm]{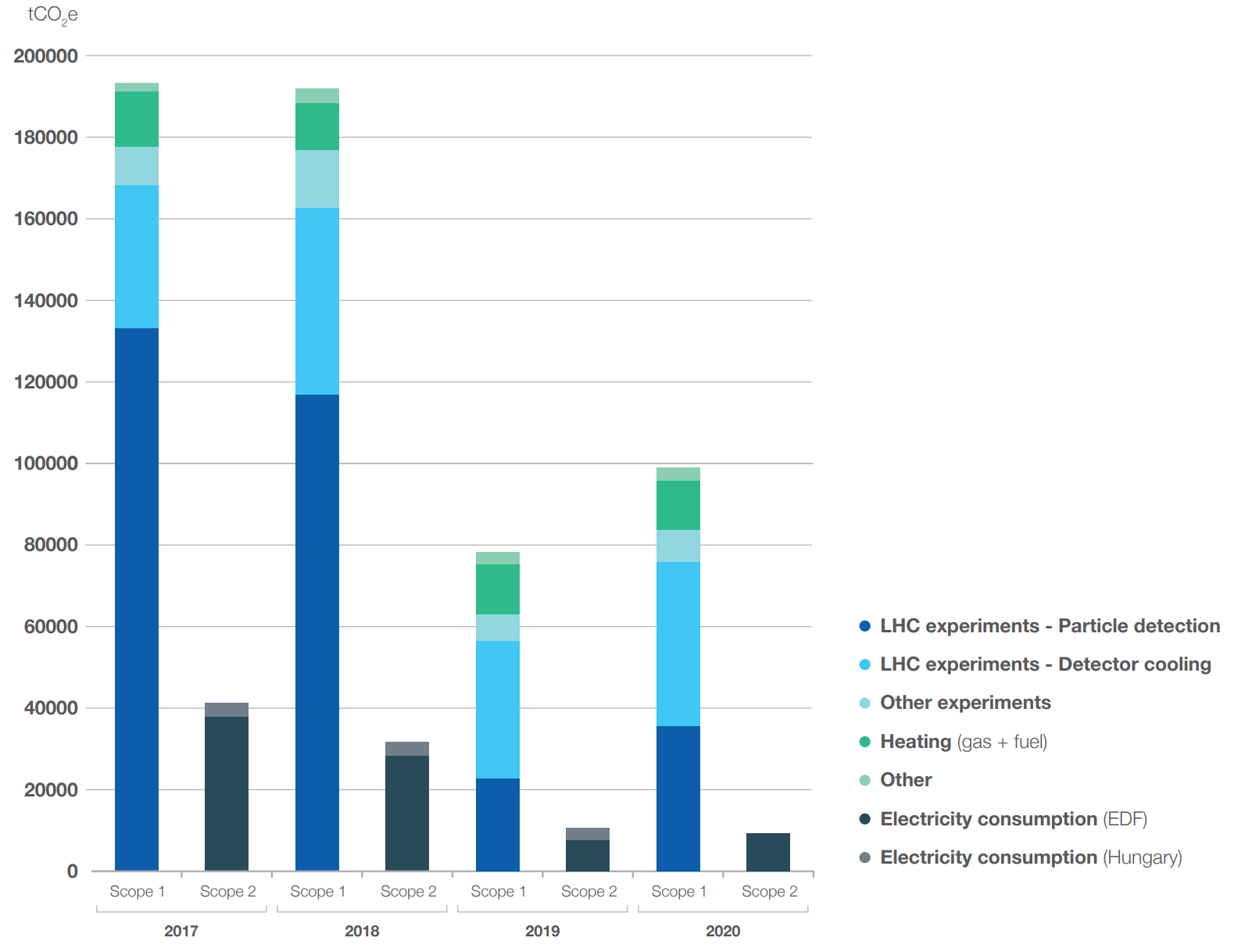}
\caption{CERN Scope 1 and 2 emissions from 2017 to 2020. The different Scopes are explained in Section~\ref{sec:activities}~\cite{bib:CERNenvurl}.}
\label{fig:CERNemissions}
\end{figure}
\begin{figure}[htbp!]
\centering
\includegraphics[width=12cm]{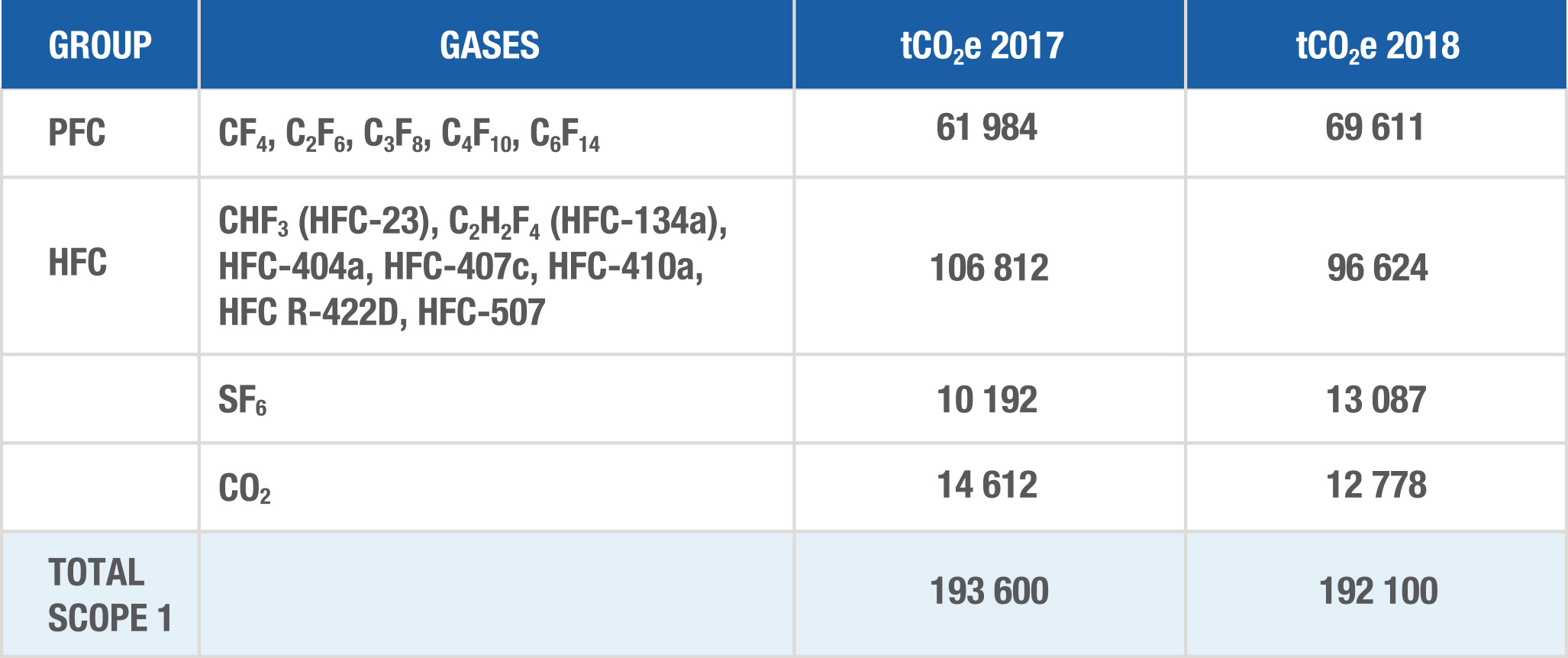}
\caption{Breakdown of Scope 1 emissions by gas types for 2017-2018~\cite{bib:CERNenv1sturl}.}
\label{fig:gas_emissionsCERN}
\end{figure}

\label{sec:gases}
According to CERN's environmental reports~\cite{bib:CERNenvurl}, the dominant source of CO$_2$e emissions at the laboratory are from greenhouse gases used for detectors and for cooling. As can be seen in Figure \ref{fig:CERNemissions}, whether the LHC experiments are in operation or in a shut down, gas emissions dominate over the emissions stemming from CERN's electricity usage. Figure \ref{fig:gas_emissionsCERN} shows the amount of emissions from various gases during 2017-18, when the LHC experiments were in operation, prior to Long Shutdown 2 (LS2). In general SF$_6$, 
hydrofluorocarbon (HFC) and perfluorocarbons (PFC) gases are used in particle detection. HFCs and PFCs are also used for detector cooling, HFCs are used in air conditioning systems, and SF$_6$ is also used for electrical insulation in power supply systems. All these gases are subject to the UN Kyoto protocol~\cite{bib:Kyoto} and their usage shall fade out according to the Kigali amendment of the Montreal protocol~\cite{bib:Kigali}. Overall, out of emissions specifically from particle detectors, 78\% of emissions come from the gas C$_2$H$_2$F$_4$ (R134a) which has a global warming potential (GWP) of 1300 times the warming potential of CO$_2$ (over a 100-year time scale)~\cite{bib:GWP}, while SF$_6$ has a GWP of 23500 and contributes 8\% of total particle detection emissions~\cite{bib:BeatriceTalk}. The types of detectors that make use of those gases include Resistive Plate Chambers (RPC), which are pairs of parallel plastic plates at an electric potential difference, separated by a gas volume. The choice of gases allows for a relatively low operating voltage, non-flammability, low cost and an adequate plateau for safe avalanche operations. Another detector is Cathode Strip Chambers, which usually includes the CF$_4$ gas (GWP of 6630). CF$_4$ gas is commonly used in particle detectors owing to its properties that prevent ageing, enhance timing resolution or because of its scintillation emission~\cite{bib:BeatriceTalk}. 

Due to their very high GWP, F-gases are under regulation in the EU~\cite{bib:EUreg} and mandatory reporting in the U.S.~\cite{bib:USreg}. Consequently, their continued procurement and price for the whole duration of the LHC program is under threat. CERN has put together various strategies to mitigate the emissions from those gases~\cite{bib:CERNenvurl}. For example, gas re-circulation is used for all gas systems across the LHC experiments. In addition, gas recuperation is also in use in some areas. During LS2, an extensive campaign of fixing leaks has occurred~\cite{bib:BeatriceTalk}. Leaks tend to be concentrated in the gas inlets, the polycarbonate gas connectors and the polyethylene pipes, so having access to the chambers is necessary to fix those leaks.

In the longer term, both for current and future detectors, finding alternative gases with lower GWP would be very beneficial, and studies are currently ongoing along those lines (e.g.~\cite{bib:gas1, bib:gas2, bib:gas3}). Although new liquids and gases have been developed for industry as refrigerants and high voltage insulating media, those are not necessarily appropriate for detector operation, especially taking into account the constraints of having to operate the current detectors. For example, there cannot be changes to the high voltage system or to the front-end electronics. Finding replacement gases needs to take into account several factors: their safety (non-flammable and low toxicity) and their environmental impact (low GWP) while maintaining their detector performance (including preventing the ageing of the detectors, ensuring good quenching and being radiation-hard)~\cite{bib:BeatriceTalk}. 

Looking to the long term future, these results highlight the crucial need to design future detectors (including cooling systems) with gas GWP in mind. For example, the proposed FCC-ee detector ideas mention using RPCs for the muon detectors and different gaseous detectors containing gas mixtures which include 10\% iC$_4$H$_{10}$ (isobutane, GWP of about 3)~\cite{bib:FCC-ee}. Over the next few years, it will be imperative to perform R\&D aimed at reducing the GHG emissions of future detectors and cooling systems as much as possible.

%% file: computing.tex
\label{sec:computing}

High-performance computing (HPC) is an essential part of physics research. It is also a growing source of greenhouse gas emissions, primarily due to the large amount of electricity used by computation itself. Across all sectors, data centers and computing already contribute approximately 2-4\% of global GHG emissions \cite{bib:lannelongue_1}, and that fraction is only predicted to grow in the next 10 years. While the environmental costs of HPC are not unknown, they are often not prioritized, or even discussed, when planning and scheduling computational projects. We believe the particle physics community should improve its accounting of this issue. We provide several specific suggestions below, many of which are inspired by Lannelongue \textit{et al.} \cite{bib:lannelongue_1}
 
The best way to reduce the carbon emissions of high-performance computing is to ensure that compute facilities are powered by carbon-free energy sources. In the early planning stages of a compute facility, when siting decisions are being made, the carbon intensity of the electricity used to power the facility should be a major factor into the decision making process. In many places, renewable electricity is already cheaper than fossil-fueled electricity, making this a cost-effective decision as well. Local installation of renewable energy generating capacity can also reduce the carbon impact of both future and existing computing facilities, most easily with rooftop solar panels. Other options may also be available depending on the location of the compute center. These strategies have a direct and immediate impact on the carbon emissions associated with a compute center, and do not rely on the cooperation of users as several of our other suggestions do, making them likely more effective.

Reductions in GHG emissions can also be achieved on the demand side with current technology, through a combination of optimization of resource use and careful planning of timing and siting of computation. These suggestions give individual users tools to directly reduce their own carbon emissions impact. However, it is often not obvious how to successfully follow them, since the information required is unavailable or difficult to access. The website \href{http://green-algorithms.org/}{Green-algorithms.org} \cite{bib:lannelongue_2} can provide a useful estimate of the cost of running a particular algorithm, but it requires knowledge of the specific hardware a computation will be using, and also the specific energy mix the compute center uses, to calculate the most accurate emissions estimate. This information could be provided by compute centers, and while the former frequently is, the latter often is not. If they aren't already doing so, compute centers should provide information on the GHG intensity of their computation, ideally in a standardized format to facilitate easy comparisons. This information could be integrated into the green-algorithms website, or else made available in a standalone website, and statistics from cloud computing services could also be included if that information is available.

Going a step further, compute centers, and the scientific collaborations that make use of them, could even provide a simple tool to estimate the GHG emissions impact of a particular job request, perhaps as a simple command-line utility. A precise estimate would be difficult to achieve, but with knowledge of the hardware, energy mix, and particulars of the job request, a reasonably accurate range of possible GHG emissions should be possible to produce, likely ranging from zero to full resource utilization. Compute centers could then report their CO2e intensity to a common location, ideally a centralized website of some sort. A version of this already exists as the Green500, an offshoot of the better-known TOP500 \cite{bib:green500}. However, the Green500 compares the power efficiency of various systems, which is related to, but not the same as, the CO2e intensity. For example, while HiPerGator AI at the University of Florida has a slightly better power efficiency rating than Perlmutter at the National Energy Research Supercomputing Center (NERSC, Berkeley, CA), 80.3\% of electricity consumption in Florida in 2019 was fossil fueled \cite{bib:fl_energy} while only 35.9\% California's electricity consumption over the same period was fossil fueled \cite{bib:ca_energy}, meaning that the per-FLOP CO2e emissions impact of Perlmutter is likely less than that of HiPerGator AI.\footnote{This is only an example to illustrate the concept; obviously the exact energy mix for each compute center is likely not identical to the statewide average.}

While obtaining better information about compute center GHG emissions can be very useful for long-term planning, such as an experiment deciding where to do the bulk of their computing, it is less useful for individual users who often do not have the ability to choose where to do their large-scale computing. Users do usually have the ability to improve the efficiency of their code, which can lead to substantial improvements in emissions if done properly. However, frequently users are writing code that makes use of large libraries, such as Geant4 or ROOT, that they do not have control over, which can limit potential improvements from optimization. Developers of these libraries should be sure to provide information on how to minimize GHG emissions, such as optimal hardware, and scaling of memory utilization. Tracking the GHG emissions efficiency over software releases with some kind of standard benchmark would provide an incentive to minimize emissions.

Even if one makes optimal choices to minimize the GHG emissions of their computation, some emissions will still inherently result from the electricity used. Even if the electricity is produced from renewable sources, the life cycle emissions of those power sources, and also of the compute center itself, are nonzero. Purchasing carbon offsets, while never a true substitute for emissions reductions, can help to mitigate the remaining unavoidable emissions. Incentivizing the purchasing of carbon offsets for unavoidable emissions is something funding agencies could do. Some Department of Energy compute centers, such as NERSC, use an allocation-and-charging framework for distributing computational resources. Carbon offsets could be folded into the cost of batch jobs by the compute center, drawn directly from the user's allocation. This would incentivize experiments to choose greener compute centers, and users to ensure that their code runs as efficiently as possible. At minimum, compute centers could provide links to legitimate offset companies, or other agencies that rate and verify them.

Finally, compute centers could coordinate their power loads with their electricity suppliers, scheduling more jobs when electricity is cheap and relatively clean (e.g. midday or nighttime), and reducing their loads when electricity is expensive and relatively dirty (e.g. late afternoon through the early evening). For the greatest impact, this would require real-time information from the electricity supplier, but if the compute center is a sufficiently large electricity customer, the supplier would likely be happy to provide this information since it would improve their demand response. Ahmed {\it et al.} have completed a study of this possibility.~\cite{bib:demand_response}

%% file: activities.tex
\label{sec:activities}

\subsection{Overview of GHG emissions from particle physics laboratories}
\label{sec:activities_fermilab}
The previous sections demonstrate that particle physics laboratories are expected to have emissions associated both with their research activities, as well as with the other work-related activities of their employees. GHG emissions fall under three scopes widely used in the reporting of emissions~\cite{bib:ghgprotocol}: Scope 1 refers to direct emissions from the organization, Scope 2 includes indirect emissions, most notably from electricity generation, heating, etc., and Scope 3 includes all other indirect emissions, upstream and downstream of the organization, including e.g. business travel, personnel commutes, catering, etc. 

\begin{figure}[htbp!]
\centering
\includegraphics[width=12cm]{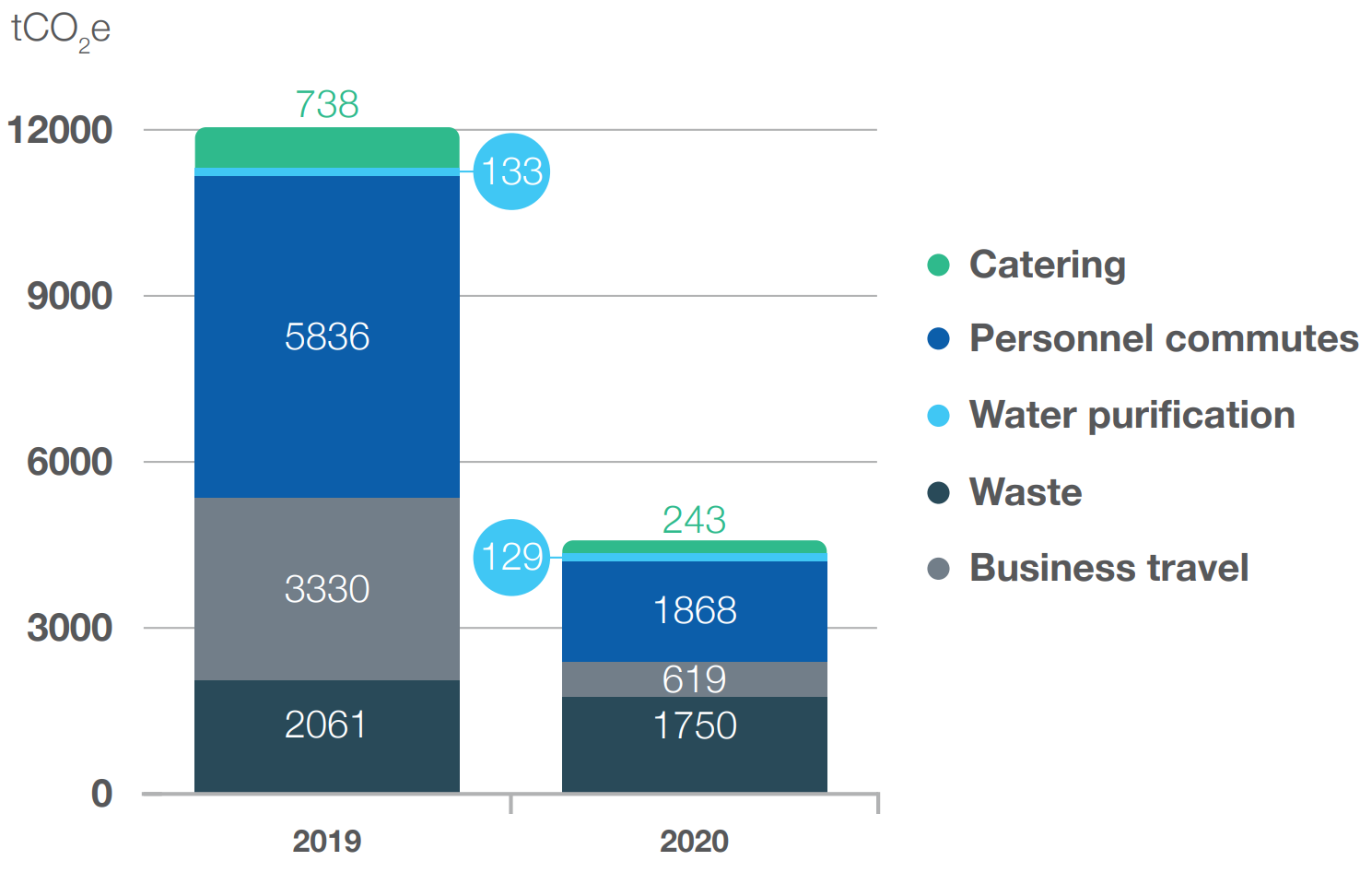}
\caption{CERN's Scope 3 emissions from 2019 and 2020~\cite{bib:CERNenvurl}.}
\label{fig:CERN_Scope3}
\end{figure}

For example, Figure~\ref{fig:CERNemissions} in Section~\ref{sec:gases} showed the emissions associated with CERN's Scope 1 and Scope 2 activities since 2017. During the time periods when the LHC is operating, the accelerator complex is the largest contributor to Scope 2 emissions. CERN obtains its electricity mainly from France and uses an emission calculation which is location-based from the ``EDF Bilans des émissions de GES". These emissions benefit from the fact that France mainly gets its electricity from nuclear power and as such it is largely carbon free (88\% in 2017)~\cite{bib:CERNenv1sturl}.  Consequently, as explained in Section \ref{sec:gases}, the largest overall emissions from CERN are from GHG gases used for detectors and cooling (part of Scope 1). In CERN's second environment report, covering the years 2019 and 2020, CERN included its first estimate of Scope 3 emissions as shown in Figure~\ref{fig:CERN_Scope3}. These do not yet include emissions from procurement, which is expected to be the largest source. CERN plans to include those in future reports. Most of the business travel emissions are from air travel, especially from long distance flights and only include the business travel of CERN employees, not from the very large number of CERN users. In a similar way, only the electricity used for on-site computing is included in the Scope 2 numbers, not the computing done on Grid sites located outside of CERN.

In the U.S, many particle physics laboratories are mandated to report on their environmental impact, including GHG emissions, every year. As an example, Fermilab's annual Site Environmental Reports~\cite{bib:fnal_env_rprts} include this information. Fermilab is committed to assist the U.S. Department of Energy in meeting reduction goals for all three scopes. Emissions are calculated based on the MRR (Greenhouse Gas Mandatory Reporting Rule~\cite{US_GHG_MRR}), and many of Fermilab’s greenhouse gas emission sources are exempt from reporting (such as emergency back-up power generators). In 2013, Fermilab was registered as a ROSS site (Registration of Smaller Sources~\cite{ROSS}), which corresponds to the adherence to certain standards and limits, e.g. limits on combined pollutants such as carbon monoxide, nitrogen oxides, sulfur dioxide, limits on mercury or lead air emissions. Table~\ref{tab:fermilab_emissions} summarizes emissions at Fermilab over the last four years and compared to the reference year of 2008, split into the different emission scopes. It has to be noted that 2020 and 2021 were highly unusual years with reduced on-site operations and occupancy during the COVID-19 pandemic. Fermilab's dominant emissions are from its electricity consumption. Fermilab is part of the PJM balancing authority, itself part of the wider Eastern Interconnection electricity grid. In 2016, the carbon intensity of PJM was 351 kg/MWh (using Emissions Factors: mass of pollutant per unit of electrical energy), which is about the average for the whole of the U.S.~\cite{bib:PJMcarbon} and, as mentioned in Section~\ref{sec:intro}, about 37\% of the national electricity grid is carbon free. Since 2011, Fermilab uses Renewable Energy Certificates (RECs) based on the lab's purchased power consumption as a primary mechanism to reduce those Scope 2 emissions. As an example, in 2013, Fermilab offset 29,000 MWh by purchasing RECs generated from new biomass plants that help reduce the environmental impact of the paper industry in Louisiana~\cite{bib:FNALToday}. In 2019 Fermilab purchased RECs to offset 11\% of its facility GHG emissions~\cite{bib:fnal_env_rprts}.

\begin{table}[htbp!]
\centering
\begin{tabular}{l|c|c|c|c|c}
 & 2008 & 2018 & 2019 & 2020 & 2021 \\ \hline
Scope 1+2 & 384,666 & 128,304 & 144,013 & 106,961 & 163,818 \\
Scope 3 & 29,503 & 16,495 & 14,468 & 6,516 & 17,456 \\ \hline
\end{tabular}
\caption{Summary of Fermilab GHG emission data from 2008 (reference year) and 2018 - 2021. Emissions are divided into the three scope areas and given in CO2e metric tons~\cite{bib:fnal_env_rprts}.}
\label{tab:fermilab_emissions}
\end{table}

\begin{figure}[htbp!]
\centering
\includegraphics[width=12cm]{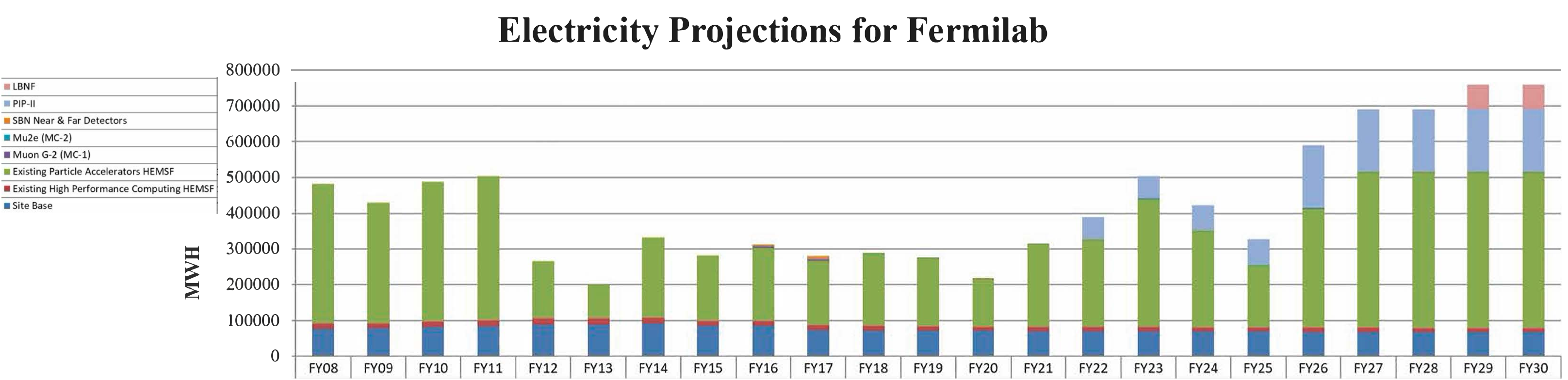}
\caption{Past and projected future electricity consumption at Fermilab split by the different accelerator projects, computing and site base operation, adapted from~\cite{bib:FNAL_sustainability}.}
\label{fig:fermilab_electricity}
\end{figure}

Past and projected electricity usage, Fermilab's major single contributor to GHG emissions, is listed in Figure~\ref{fig:fermilab_electricity}. Significant changes to Fermilab’s projected energy consumption are anticipated beginning 2024, when a planned 2-year accelerator shutdown begins to bring the new PIP-II accelerator complex online. During the shutdown, the Long Baseline Neutrino Facility infrastructure will tie into the existing accelerator complex. The addition of the PIP-II accelerator and beam requirements for LBNF are expected to increase Fermilab’s overall energy consumption by 30\% over historic peak levels.

\subsection{Overview of GHG emissions from the academic system}
Numerous activities in the academic system result in GHG emissions, from campus operations at universities, to construction and operation of large-scale research infrastructure, to academic travel for various activities such as conferences, committee meetings, and funding reviews. To provide an impression of the amount of GHG associated to these aspects of academia, we discuss several examples of activities and compare their associated GHG emissions to the personal budget of 1.1 t per year mentioned in Section \ref{sec:intro}.  

In Europe, GHG emissions from campus operations can amount to 33-35~t CO2e emissions per employee per year, of which more than 75\% can be Scope 3 emissions, including air travel and the supply chain (including consumables and equipment for research)~\cite{bib:Borgermann}. The estimates of many universities are significantly lower than this, because Scope 3 emissions are not comprehensively accounted for. For example, ETH Z\"urich quotes 2.8 (3.0) tCO2e per FTE employee in 2012 (2018) but notes that data for Scope 3 is only available for selected domains~\cite{bib:ETH}.
 In the U.S., numbers tend to be higher. For instance, MIT reports about 193 kt CO2e per year as on-campus emissions, which amounts to about about 12 t CO2e per employee~\cite{bib:MIT}. This is exclusive of all Scope 3 emissions. At UCLA, Scope 1 and 2 emissions together with commuting and business travel, i.e., with a very incomplete accounting of Scope 3, come to about 7 t CO2e per year per full-time-equivalent ~\cite{bib:UCLA}. At academic institutions, there is significant heterogeneity in the reporting of GHG emissions from flying \cite{bib:KreilStauffacher2021}. For example, often only trips paid by the institutions are accounted for, and so trips on external grants are not included in any institution's accounting.

%% the below text is now in the previous section essentially
%Among research institutes in particle physics, GHG emissions have been estimated for CERN and amount to about 190~kt CO2e per year (before the current shutdown)~\cite{bib:CERN}. This does not account for the supply chain nor for air travel by CERN users who are not CERN employees\footnote{At academic institutions, there is significant heterogeneity in the reporting of GHG emissions from flying \cite{bib:KreilStauffacher2021}. Often, only those trips are accounted for, for which the institution pays, such that, e.g., trips on external grants are not included in any institution's accounting.}. Similarly, electricity for computing by CERN users is not included, if it is done off-site, as only the on-site electricity consumption is included. Thus, the estimate is a low one, with the full amount of emissions likely significantly higher (judging from the the impact that a comprehensive accounting of Scope 3 emissions has on universities' carbon budgets). Per CERN employee, the per-capita emissions (with the LHC running) are 76~t CO2e; per CERN user 16~t (https://home.cern/about/who-we-are/our-people). 

Similar numbers arise in the related research field of astronomy, which also relies on large-scale infrastructure as well as computing. Jahnke~\cite{bib:Jahnke} estimates about 18~t CO2e emissions per year per researcher at the Max-Planck-Institute for Astronomy in Germany, with 8.5~t for air travel and roughly 5~t for computing. In Australia (see discussion in \cite{bib:Jahnke}, based on \cite{bib:Stevens}), where less electricity comes from renewable sources and more air miles are flown, the total emissions are more than twice as high.

%The data from CERN, which does not comprehensively account for air travel nor for computing by particle physicists, together with the data from astronomy, which accounts for both factors, suggest that there is an urgent need to reduce research-related GHG emissions in particle physics.

It is worth considering the structure of future collaborations and requirements to be at the experimental site in person. As air travel has become cheaper and funds to bring along family members for extended stays to particle physics laboratories have become considerably smaller, regular commutes of working group conveners to the labs have become the new normal. This is particularly the case for positions in the LHC collaborations that are expected to have ``a significant presence at CERN". While this has been largely waived during the COVID pandemic, it is not yet clear how this policy will evolve in future. Similarly, there are detector operations and data taking shifts that might be moved to remote locations. A number of institutes created remote control rooms before the start of the LHC, but the extent to which they have been used and the emissions saved is  undocumented. Generally, as the size of the collaborations is expected to grow in size and in geographical diversity, regional centers may be established as alternative hubs to the main experimental site, as Fermilab can already be considered today for CMS. If successful, such a system might reduce the travel-related carbon emissions by more than half. For example, according to the carbon-offset provider \href{https://www.atmosfair.de/en/}{atmosfair}, a return flight from San Francisco to Geneva has four times the carbon emissions compared to a return flight to Chicago only. 

There are additional sources of GHG emissions in the academic system which are also relevant in particle physics. 
One example is travel for committee meetings, funding reviews and the like. Data on this is scarce, but an example exists from the European Research Council~\cite{bib:Bousema}, which estimates more than 1 t CO2e emissions per review, due to travel of the international panel as well as the international pool of applicants.  Similar examples could be found in the U.S., such as funding agency reviews of construction projects or operations programs.  When these reviews are held in person, at least 25 people need to travel from around the country to a single location, which is typically not chosen with a goal of minimizing climate impacts.

Finally, conference travel is a much-debated source of GHG emissions in the academic system \cite{bib:Bjorkdahl}, and the term ``conference tourism" \cite{bib:Hoyer} has been coined, relating to conferences at exotic locations.  Across various disciplines, the systematic study in~\cite{bib:Spinellis} estimates that typically about 1~t CO2e are emitted per participant, and calculations from various individual examples result in similar average emissions \cite{bib:Jaeckle,bib:Burtscher}. Reviews of the International Conference on High Energy Physics (ICHEP) estimate about 2~t CO2e per participant using the methodology employed in~\cite{bib:agu}, which accounts not only for CO$_2$ emissions but also for the impact due to contrails, ozone formation and other effects. The notable exception was the 2012 edition in Melbourne, causing per participant emissions of about four times the average due to its remote location, highlighting to some extent the tension between geographical inclusion and reducing travel emissions.

Likely more relevant than the average emissions is the distribution of emissions. Often, a smaller fraction of participants produces the largest share of emissions, e.g., in~\cite{bib:Burtscher}, 10\% of the flights taken resulted in over 50\% of the emissions. Thus, long-haul flights are a particular challenge. Reductions of anywhere between 94\% and 98\% of CO2e emissions have been estimated for fully virtual meetings. Similarly, hub-based formats, which eliminate long-haul travel by collecting participants at regional hubs and link those hubs virtually, are expected to result in significant reductions \cite{bib:agu}. When applied to ICHEP, the emissions could be reduced to 15-35\% of those of a traditional one-hub ICHEP conference.

Even for fully in-person meetings, reductions of 20\% of GHG emissions have been found by optimizing the meeting location~\cite{bib:Stroud};  with some even finding variations of up to a factor of two~\cite{bib:Jaeckle}.

Calls for the reduction (not necessarily complete elimination) of air travel to conferences are often met with the worry as to whether this will result in a reduced amount of networking, dissemination of results and similar. It has long been assumed that there is a direct correlation (or even causation) between the amount of air travel by a scientist and their scientific impact. Quantitative studies are now starting to question this correlation.  \cite{bib:Wynes} and \cite{bib:Ciers} do not find a correlation between the amount of air travel and the h-index in their studies of researchers across multiple fields. 

Nevertheless, a reduction of air travel in science would have to be implemented carefully, in order to avoid unintended consequences. This could include early-career researchers finding it more challenging to build their networks, or that researchers from the Global South, who typically have to travel further to meet colleagues or attend conferences, are cut off from scientific exchanges. Both can be avoided by carefully evaluating for which purposes and by whom air travel may be scientifically valuable, when in other cases it could be substituted by a virtual exchange.

Moreover, holding meetings in hybrid or virtual form can have important co-benefits, in particular for inclusivity, strengthening the case for hybrid/virtual meetings.
For instance \cite{bib:Sarabipour} shows a significant increase in participation from countries outside North America, including in Latin America and Africa, when the APS April meeting shifted to online format. 
Further, \cite{bib:Skiles} found an increase in female participation (and abstract submission from female researchers), when conferences shifted from in-person to online format.
Climate sustainability and (global) inclusivity can thus be achieved together.
 
Additionally, GHG emissions at conferences are related to accommodation as well as catering. The effect of offering fully vegetarian/vegan catering during a four-day conferences was estimated in \cite{bib:Jaeckle} to result in about 16 (vegetarian) or 23 (vegan) t CO2e emissions for a conference series with an average of about 1600 participants. For comparison, the GHG emissions from air travel, which depended on the choice of location, varied between 900 and 1800 t CO2e. 

%% file: recs.tex
\label{sec:recs}
We offer the following recommendations on how we as a field can reduce our impact on the climate and moderate the ongoing trends of global climate change.
\begin{itemize}
    \item New experiments and facility construction projects should {\bf report on their planned emissions and energy usage as part of their environmental assessment}, which will be part of their evaluation criteria.  These reports should be inclusive of all aspects of activities, including construction, detector operations, computing, and researcher activities.
    
    \item U.S. laboratories should be involved in a {\bf review across all international laboratories to ascertain whether emissions are reported clearly and in a standardized way}. This will also allow other U.S. particle physics research centers (including universities) to use those standards for calculating their emissions across all scopes.
    
    \item Using the reported information as a guide, all participants in particle physics -- laboratories, experiments, universities, and individual researchers -- should {\bf take steps to mitigate their impact on climate change by setting concrete reduction goals and defining pathways to reaching them} by means of an open and transparent process involving all relevant members of the community.  This may include {\bf spending a portion of research time on directly tackling challenges related to climate change in the context of particle physics}.
    
    % following one has been merged with above
%    \item Management of experiments and facilities should set concrete reduction goals and define pathways to reaching them by means of an open and transparent process involving all relevant members of the community. Accountability should be ensured, if goals are not reached.
    
    \item U.S. laboratories should invest in the development and affordable deployment of next-generation digital meeting spaces in order to {\bf minimize the travel emissions of their users}. Moreover the particle physics community should actively promote hybrid or virtual research meetings and travel should be more fairly distributed between junior and senior members of the community. For in-person meetings, the meeting location should be chosen carefully such as to minimize the number of long-distance flights and avoid layovers.

%    \item New facility construction projects should report on their planned emissions as part of their environmental assessment, which will then serve as part of the assessment of any proposal for a facility.  Among other things, this will require standardized methods of calculating carbon impacts.
    \item  Long-term projects should {\bf consider the evolving social and economic context}, such as the expectation of de-carbonized electricity production by 2040, and the possibility of carbon pricing that will have an impact on total project costs.
   % \item Research and development efforts on civil construction should include investments in mitigating climate impacts through conscious choices of construction materials and processes, including the potential use of carbon offsetting mechanisms.
    \item All U.S. particle physics researchers should {\bf actively engage in learning about the climate emergency and about the climate impact of particle-physics research.} 
    
    \item The U.S. particle physics community should {\bf promote and publicize their actions surrounding the climate emergency to the general public and other scientific communities}.
   % \item U.S. particle physics researchers should be encouraged to spend a fraction of their research time on directly tackling challenges related with the climate emergency.
    \item The U.S. particle physics community and funding agencies should {\bf engage with the broader international community to collectively reduce emissions}.
\end{itemize}

%% file: Ack.tex
The authors would like to thank many colleagues for crucial input and discussions and in particular Roberto Guida, Johannes Gutleber, Sonja Kleiner, Karen Kosky, Beatrice Mandelli, Eric Mieland, and Marlene Turner.

%% file: references.tex
% BibTeX or Biber users please use (the style is already called in the class, ensure that the "woc.bst" style is in your local directory)
% \bibliography{name or your bibliography database}
%
% Non-BibTeX users please use
%